\documentclass[aps,pre,twocolumn,superscriptaddress,reprint]{revtex4-2}

\usepackage{amsmath}
\usepackage{amsfonts}
\usepackage{graphicx}
\usepackage{cancel}

\usepackage[colorinlistoftodos,prependcaption,textsize=small]{todonotes}
\usepackage{color}
\definecolor{red}{rgb}{0.75,0,0}
\definecolor{blue}{rgb}{0,0,0.75}
\definecolor{green}{rgb}{0,0.5,0}
\newcommand{\yzh}[1]{{\color{black} #1}}

\newcommand{\mcm}[1]{{\color{black} #1}}
\newcommand{\mcmrev}[1]{{\color{black} #1}}
\newcommand{\rev}[1]{{\color{black} #1}}

\makeatletter

\newcommand{\Hb}{\mathbf{H}}
\newcommand{\Qb}{\mathbf{Q}}

\newcommand{\vb}{\mathbf{v}}

\newcommand{\sigb}{\boldsymbol{\sigma}}
\newcommand{\nabb}{\boldsymbol{\nabla}}

\newcommand{\eq}{\begin{equation}}
\newcommand{\eeq}{\end{equation}}
\newcommand{\aeq}{\begin{equation}\begin{aligned}}
\newcommand{\eaeq}{\end{aligned}\end{equation}}

\makeatother

\begin{document}
\title{Vorticity phase separation and defect lattices in the isotropic phase of active liquid crystals}

\author{Fernando Caballero}
\thanks{These authors contributed equally}
\affiliation{Department of Physics, University of California Santa Barbara, Santa Barbara, CA 93106, USA}

\author{Zhihong You} 
\thanks{These authors contributed equally}
\affiliation{Fujian Provincial Key Laboratory for Soft Functional Materials Research, Research Institute for Biomimetics and Soft Matter, Department of Physics, Xiamen University, Xiamen, Fujian 361005, China}
 
\author{M. Cristina Marchetti}
\affiliation{Department of Physics, University of California Santa Barbara, Santa Barbara, CA 93106, USA}

\date{\today}
	
\begin{abstract}
We use numerical simulations and linear stability analysis to study the dynamics of an active liquid crystal film on a substrate in the regime where the passive system would be isotropic. Extensile activity builds up local orientational order and  destabilizes the quiescent isotropic state above a critical activity value, eventually resulting in spatiotemporal chaotic dynamics akin to the one observed ubiquitously in the nematic state. Here we show that tuning substrate friction yields a variety of emergent structures at intermediate activity, including lattices of flow vortices with associated regular arrangements of topological defects and a new state where flow vortices trap pairs of $+1/2$ defect that chase each other tail. \mcmrev{These chiral units spontaneously pick the sense of rotation and organize in a hexagonal lattice, surrounded by a diffuse flow of opposite rotation to maintain zero net vorticity.} The length scale of these emergent structures is set by the screening length $l_\eta=\sqrt{\eta/\Gamma}$ of the flow, controlled by the shear viscosity  $\eta$ and the substrate friction $\Gamma$, and  
can be captured by simple mode selection of the vortical flows.
We demonstrate that the emergence of coherent structures can be interpreted as a phase separation of vorticity, where friction plays a role akin to that of birth/death processes in breaking conservation of the phase separating species and selecting a characteristic scale for the patterns. Our work shows that friction provides an experimentally accessible tuning parameter for designing controlled active flows.
\end{abstract}
	
\maketitle

\section{Introduction}

The unique way in which active liquid crystals transform energy into directed motion is responsible for a number of phenomena not present in equilibrium, such as self-sustained laminar flows~\cite{aditi2002,marchetti2012soft} and spatio-temporal chaotic flows 
known as active turbulence~\cite{alert2022active}.


Active turbulence is observed in a variety of systems, from liquid crystalline fluids reconstituted from cell extracts to epithelial monolayers.  It has been quantified in active liquid crystals of cytoskeletal microtubule bundles cross-linked by kinesin, a motor protein that consumes ATP as it moves along the bundles, creating extensile stresses on the flow in which they are submerged \cite{lemma2021multiscale,tayar2021active,tayar2022controlling,sanchez2011cilia,Adkins2022}. Active liquid crystals have been studied extensively through continuum theories, and there is now a large body of analytical and numerical work  characterizing their stability, dynamical regimes  and interactions with other immiscible species~\cite{giomi2011,giomi2012,doostmohammadi2017,shendruk2017,opathalage2019,hardouin2019, kempf2019active, caballero2022activity}.

Controlling this chaotic spontaneous flow to create coherent structures holds the promise of engineering active fluids for microfluidic applications and functional materials capable of delivering directed mechanical forces~\cite{bowick2022symmetry}. Active liquid crystals have become  promising candidates for these applications~\cite{Adkins2022,tayar2022controlling}, as 
physical confinement, substrate friction and substrate patterning
allow 
control of \mcm{chaotic flow 
and development of coherent structures.} 
These effects have been mainly examined so far 
in \mcm{regimes of parameters corresponding to the ordered nematic state of the passive} liquid crystal \cite{doostmohammadi2016,guillamat2016,guillamat2017,thijssen2020,martinez2021scaling,thijssen2020role,thampi2014active}.

In this work, we \mcm{examine numerically the effect of substrate friction on two-dimensional active nematic liquid crystals in a regime of parameters where the fluid is isotropic when passive} 
and we reveal \mcm{the emergence of new coherent structures.} 
\mcm{These include previously observed~\cite{doostmohammadi2016,chandragiri2020flow,keogh2022helical} lattices of flow vortices with an associated regular arrangement of half integer disclinations and a novel state where flow vortices trap pairs of $+1/2$ defects forcing them to rotate and chase each other's tail. \mcmrev{This state breaks the symmetry between positive and negative vorticity, as the $+1/2$  pairs are all trapped in vortices of the same sign and coherently rotate in the same direction. The requirement of zero net vorticity is mantained by diffuse counterrottaing flows that permeate the interstitial region between localized vortices.} Similar configurations \mcmrev{of $+1/2$ pairs chasing each other's tails} have been observed in active nematic trapped in small circular wells~\cite{opathalage2019}, \mcmrev{where the required net topological charge is $+1$,} \mcmrev{or in system with  anisotropic and spatially varying friction~\cite{thijssen2021submersed, thijssen2020active}}. Here, in contrast, both the \mcmrev{vortex lattice} and the \mcmrev{the coherent chiral motion of defect pairs} emerge spontaneously in unconfined bulk fluids.} These structures arise \mcm{in a narrow range of activity} from the interplay between 
the \mcm{tendency of the unbound system to destabilize on length scales $\sim|\alpha|^{-1/2}$, \rev{(where $\alpha$ is the \mcmrev{scale of the active stress,} }
and the stabilizing effect of \mcmrev{substrate friction $\Gamma$} that screens flows on scales of the order of the viscous length $l_\eta=\sqrt{\Gamma/\eta}$, \mcmrev{with  $\eta$ the shear viscosity}. This length control\mcmrev{s} the size of the flow structures in this intermediate activity regime. }
\mcm{The numerical work is complemented by linear stability analysis that identifies the length scales of the emergent structures, in excellent agreement with numerics. Finally, we show that a simple ansatz for the lattice-like structure of flow vortices naturally reproduces the associated defective nematic texture, demonstrating that the established connection between flow and defect structures in these active liquid crystals holds in the regime where the passive system is isotropic.  }

\begin{figure*}[t]
\includegraphics[width=1\textwidth]{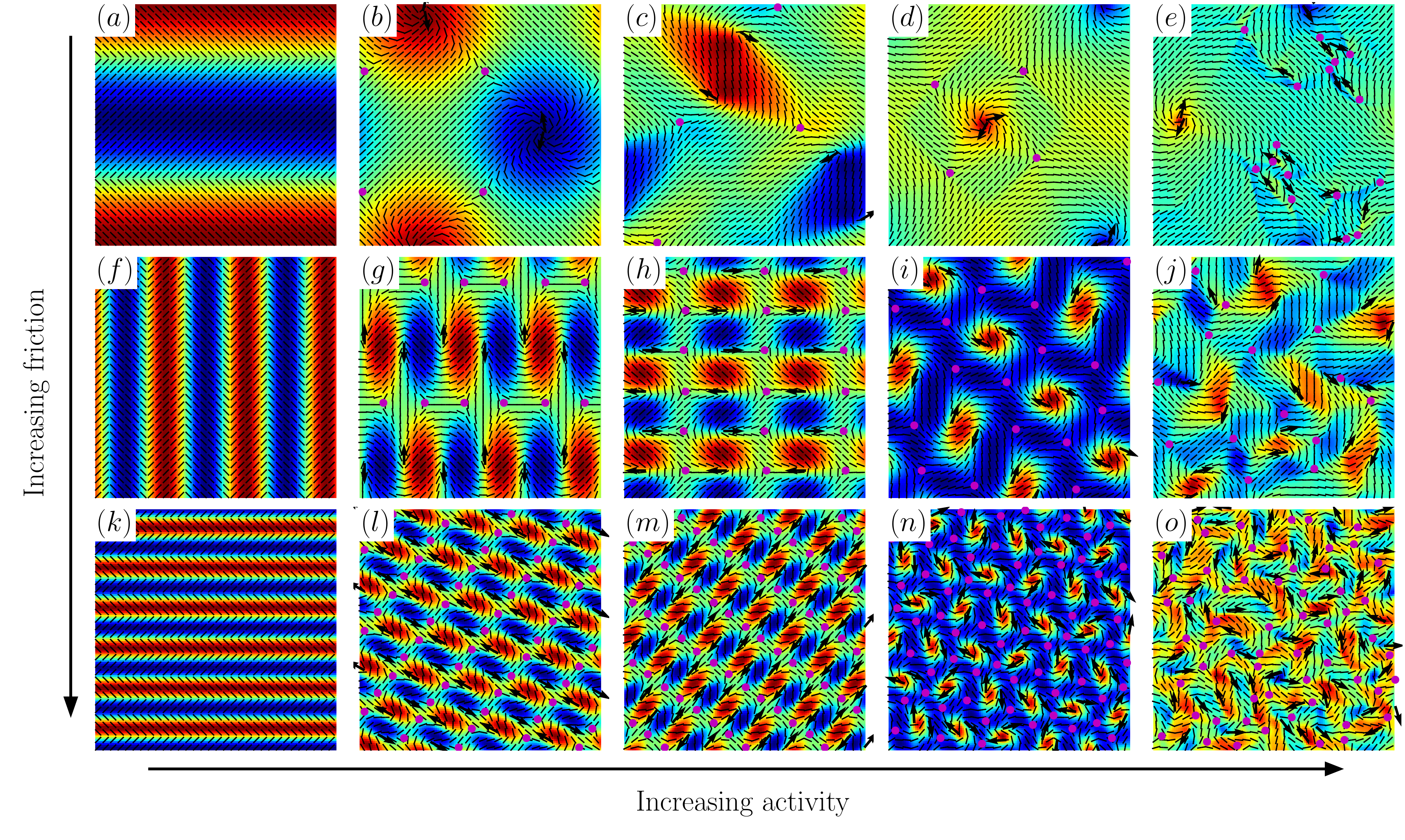}
\caption{\label{fig:stat-diag} Structures observed by increasing substrate friction $\Gamma$ (top to bottom) and activity $\alpha$ (left to right respectively), as indicated by the arrows. The color represents the vorticity, the lines indicate the nematic director, the dots and arrows indicate, respectively, $-1/2$ and $+1/2$ defects, with the arrows pointing in the direction of the polarization of the $+1/2$ defects. Colorbars have not been added to avoid clutter, since each frame has a different vorticity. The first row shows \mcm{the behavior for $\Gamma = 0$, where the scale of the observed structures is controlled by the} system size, for $|\alpha|=3.6, 3.62, 5, 8, 50$. The second row corresponds to $\Gamma=0.01$ and $|\alpha|=4.23, 4.5, 4.8, 5, 8$. The third row is for $\Gamma=0.15$ and $|\alpha|=6.46, 6.5, 8, 10, 15$. \mcm{Defects are denoted by red arrows ($+1.2$) and magenta dots ($-1/2$).} As we increase activity, the system transitions from a uniform state to a band state (first column), then vortex lattices (columns 2-4) and finally to states of active turbulence (column 5).}
\end{figure*}

\mcm{One can draw a suggestive} parallel between the emergence of coherent structures observed here and \mcm{motility induced} phase phase separation (MIPS) \mcmrev{in the absence of number conservation~\cite{cates2010}}.
\mcm{In MIPS a conserved density of Active Brownian Particles spontaneously undergoes bulk phase separation into a dense liquid and an active gas~\cite{tailleur2008statistical,fily2012athermal,redner2013structure,cates2015motility}. Breaking mass conservation arrests phase separation and} stabilizes low wavelength modes, \mcm{yielding regular arrays of dense droplets or rings with a characteristic steady state length scale~\cite{cates2010}. Similarly, our system can be thought of as undergoing phase separation into regions with opposite}
sign of vorticity. In absence of substrate friction, 
\mcm{when momentum is conserved, vortical phase separation spans the entire system and the scale of the vortical flows is set by the}
system size, \mcm{as shown in Fig.~\ref{fig:stat-diag}(b).}
Substrate friction \mcm{breaks momentum conservation} and 
screens flows 
on scales $l_\eta$, 
effectively arresting vortical phase separation.
The result are ordered micro phase-separated states 
with rectangular and distorted hexagonal lattices of flow vortices,  in which pairs of $+1/2$ defects chase each other, as hinted 
previously
with effective theories of active flows \cite{slomka2017geometry}. 

In the remainder of this paper, we first introduce the model we use to get all of our results, following previous continuum theories of active liquid crystals \cite{marchetti2012soft}. We then show that a linear stability analysis properly captures the behaviour of the system as the isotropic state becomes unstable due to active flows. We show that the length scales predicted by this analysis for the emerging structures are in excellent agreement with the lengthscales we observe numerically. Finally, we will show that the lattice structures for the vorticity can be captured following previous work on nonequilibrium hydrodynamics, by which we can construct stream profiles that are static solutions to the linear Stokes' flow, and properly reproduce the observed vorticity profiles, as well as the texture of the liquid crystal in said lattice states.
\section{Hydrodynamic Model}
	
We consider a familiar model for a two-dimensional active liquid crystal  on a substrate.  The state of the system is described in terms of  the velocity field, $\textbf{v}$, and the nematic tensor, $\textbf{Q}\equiv S(\textbf{n}\textbf{n}-\textbf{I}/2)$. Here, the director $\textbf{n}$ is a unit vector identifying the direction of order  and $S$ is the nematic order parameter, with 
$\textbf{I}$ the identity tensor. The dynamics is governed by the following equations, \rev{used before to describe active liquid crystals with substrate friction \cite{thampi2014active}}
\begin{subequations}
  \begin{align}
  \label{eq:dtQ}
  D_{t}\Qb = &\lambda\textbf{D}+\Qb\cdot\boldsymbol{\omega}-\boldsymbol{\omega}\cdot\Qb+ \gamma^{-1}\Hb,\\
  \label{eq:dtv}
  \rho D_{t}\vb = &\eta\nabb^2\vb-\nabb P-\Gamma \vb+ \nabb\cdot\sigb,
  \end{align}
\end{subequations}
where $D_{t}=\partial_{t}+\vb\cdot\nabb$ is the material derivative. The first term on the right hand side of Eq. \eqref{eq:dtQ} tends to align the nematic director with the local strain rate $\textbf{D}=(\nabb\vb+\nabb\vb^{T})/2$, with $\lambda$ the flow alignment parameter. The second and third terms capture co-rotation of the director with the local vorticity $\boldsymbol{\omega}=(\nabb\vb-\nabb\vb^{T})/2$. The last term describes relaxation to minimize the Landau-de Gennes free energy,
\begin{equation}
F_{LdG}=\int_{\textbf{r}}\frac{a}{2}\text{Tr}(\Qb^2)+\frac{b}{4}\text{Tr}(\Qb^2)^2+\frac{K}{2}(\partial_iQ_{jk})^2,
\end{equation}
with $\Hb=-\delta F_{LdG}/\delta \Qb$ and $\gamma$ a rotational viscosity. 
 The free energy captures an order-disorder  transition upon tuning the parameter $a$, with $b>0$.  For $a>0$ the ground state is  an isotropic fluid with $S=0$. For $a<0$ the equilibriumn state is nematic with $S=\sqrt{-2a/b}$. Finally, $K$ is a stiffness constant that characterizes nematic elasticity, assumed for simplicity to be isotropic.

The velocity is governed by the Navier-Stokes equation,  Eq. \eqref{eq:dtv}, with viscosity $\eta$, density $\rho$ and the condition of incompressibility $\nabb\cdot\vb=0$ that determines the pressure $P$. The third term on the right hand side of Eq. \eqref{eq:dtv} is the frictional force from the substrate, with $\Gamma$ a friction per unit area. Finally, the liquid crystalline degrees of freedom create a stress $\sigb=\sigb^{e}+\sigb^{a}$ on the flow that includes the elastic stress 
\begin{equation}
\sigb^e=-\lambda\Hb+\left(\Qb\cdot \Hb-\Hb\cdot\Qb\right)\;,
\end{equation}
and the active stress $\sigb^{a}=\alpha\Qb$ that describes the effect of active force dipoles on the fluid.
Here we consider the case of extensile active stresses, corresponding to $\alpha<0$, as appropriate for instance for microtubule-kinesin suspensions.

 We focus below on the case $a>0$, which corresponds to the situation where the passive liquid crystal is in the isotropic state.  We rescale lengths with the nematic correlation length $\ell_{c}=\sqrt{K/a}$, times with the nematic correlation time $\tau_{c}=\gamma/a$, and stress with the typical nematic relaxational stress $a$. Unless otherwise specified, all \mcm{numerical} results are presented in dimensionless units. We use the following values of parameters: $\rho=0.04$, $\eta=1$, $\lambda=0.7$, $b=1000$, and vary $\alpha$, and $\Gamma$. 
 We have observed that the coherent structures described below  occur in the isotropic regime of the passive liquid crystal and are most easily observed close 
 to the critical point \rev{related to the isotropic/nematic transition, i.e. $a=0$}. This can also achieved by setting $b\sim\mathcal{O}(1)$ and choosing  a small positive value of $a$.
 Choosing $b\gg a$ overdamps the active liquid crystal, but does not increase the correlation length $l_c$, which should diverge at the critical point. To observe the coherent structures reported here, $l_c$ must remain smaller than the screening length $l_\eta$, which sets the length scale of the structures themselves. We have observed that the interval of activities in which we observe these coherent structures becomes wider if the system is placed closer to the critical point by increasing $l_c$ (for instance by increasing $K$) without making it greater than $l_\eta$.  The structures can, however, also be observed away from the critical point.
 
 We have integrated numerically Eqs. \eqref{eq:dtQ} and \eqref{eq:dtv} in a periodic square box of size $L\times L$. The default system size is $L=64$, but we have also investigated other values of $L$. Finite difference is used to discretize the system on a uniform square grid with a grid size of $0.5$. To integrate in time, we use the Runge–Kutta–Chebyshev scheme, which provides enhanced numerical stability and allows for a large time step $\Delta t=0.1$. All simulations are initialized with zero velocity and a nearly zero nematic tensor with a  small  random perturbation.

\section{Spatio-temporal patterns induced by activity: Numerical results}

The variety of spatiotemporal structures  obtained upon varying activity and  substrate friction is shown in Fig.~\ref{fig:stat-diag}. At low activity (not shown in the figure), the system is homogeneous with no flow and zero nematic order ($S=0$). 
Increasing activity destabilizes the isotropic state and drives local nematic order that eventually organizes in a variety of coherent structures.  Some of these structures, specifically the rectangular lattices in the second column of figure \ref{fig:stat-diag}, have been observed before in active liquid crystals coupled to phase separating fields \cite{doostmohammadi2016}. We show here that these regular lattices of flow vortices and nematic texture are also found without coupling to additional fields, but rather emerge spontaneously from the interplay of active stresses and flow screening.

\begin{figure}[t]
\includegraphics[width=0.45\textwidth]{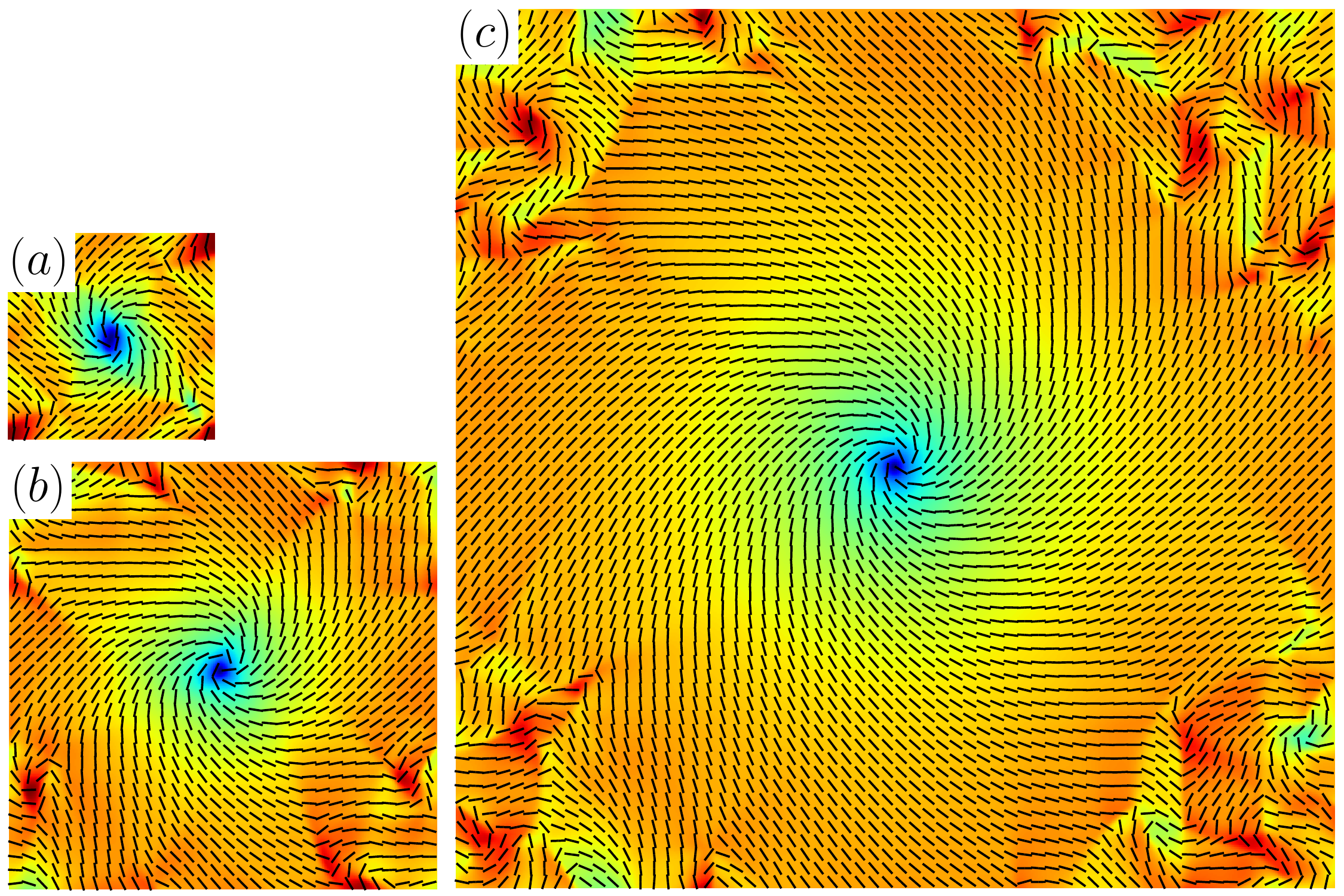}
\caption{\label{fig:SS-Ls} The steady state flow structures for $\Gamma=0$ and different system sizes: (a) L=64, (b) L=128, (c) L=256 show that in absence of friction the scale of emergent structures is controlled by the system size. \rev{All parameters except system size are the same across all three frames.} The color indicates the vorticity and the lines indicate the orientation of the nematic director.}
\end{figure}

For vanishing substrate friction, flow structures evolve with increasing activity in a manner similar to what observed in the well-studied nematic state, as shown in the top row of Fig.~\ref{fig:stat-diag} and studied before~\cite{srivastava2016,putzig2016instabilities,oza2016antipolar,santhosh2020}. The patterns are controlled by the interplay of the system size $L$ and the active length $l_\alpha\sim\sqrt{K/|\alpha|}$.  Just above the critical activity for the instability of the isotropic quiescent state, where $l_\alpha\gg L$, we observe system-spanning  structures consisting of  two parallel bands, with opposite flow directions and a zigzag nematic orientation (Fig. \ref{fig:stat-diag}(a) ~\cite{srivastava2016,Vafa2021,doostmohammadi2016}. At higher activity, the bands are replaced by two system-spanning vortices, separated by  large regions of nematic order interrupted by pairs of topological defects. 
Figure~\ref{fig:SS-Ls} highlights that in this regime flow structures are indeed controlled by 
the system size, scaling up to span the whole system  for $L$ up to
$L=256$. 

\rev{Surprisingly, this state displays stable $+1$ defects. While  $+1$ defects \mcmrev{are generally unstable at finite activity and decay}  into pairs of $+1/2$ defects, we observe here that there is a range of activity where the nematic order is enslaved to the flow and 
$+1$ defects \mcmrev{can be trapped}  at the center of \mcmrev{flow} vortices, as seen in Fig \ref{fig:SS-Ls}.}

Further increasing activity yields $l_\alpha\sim L$ and promotes defect pair unbinding. The defects unbind in the large shear rate regions at the boundaries between opposite vorticity and organize in lanes that slide past each other 
(Fig. \ref{fig:stat-diag}(c,d) and SI video 1). 
Further increasing activity, yields $l_\alpha\ll L$.
Unbound defect pairs then proliferate, rendering the system's dynamics chaotic and leading it into active turbulence (Fig. \ref{fig:stat-diag}(e) and SI videos 2 and 3).


\begin{figure}[t]
\includegraphics[width=0.5\textwidth]{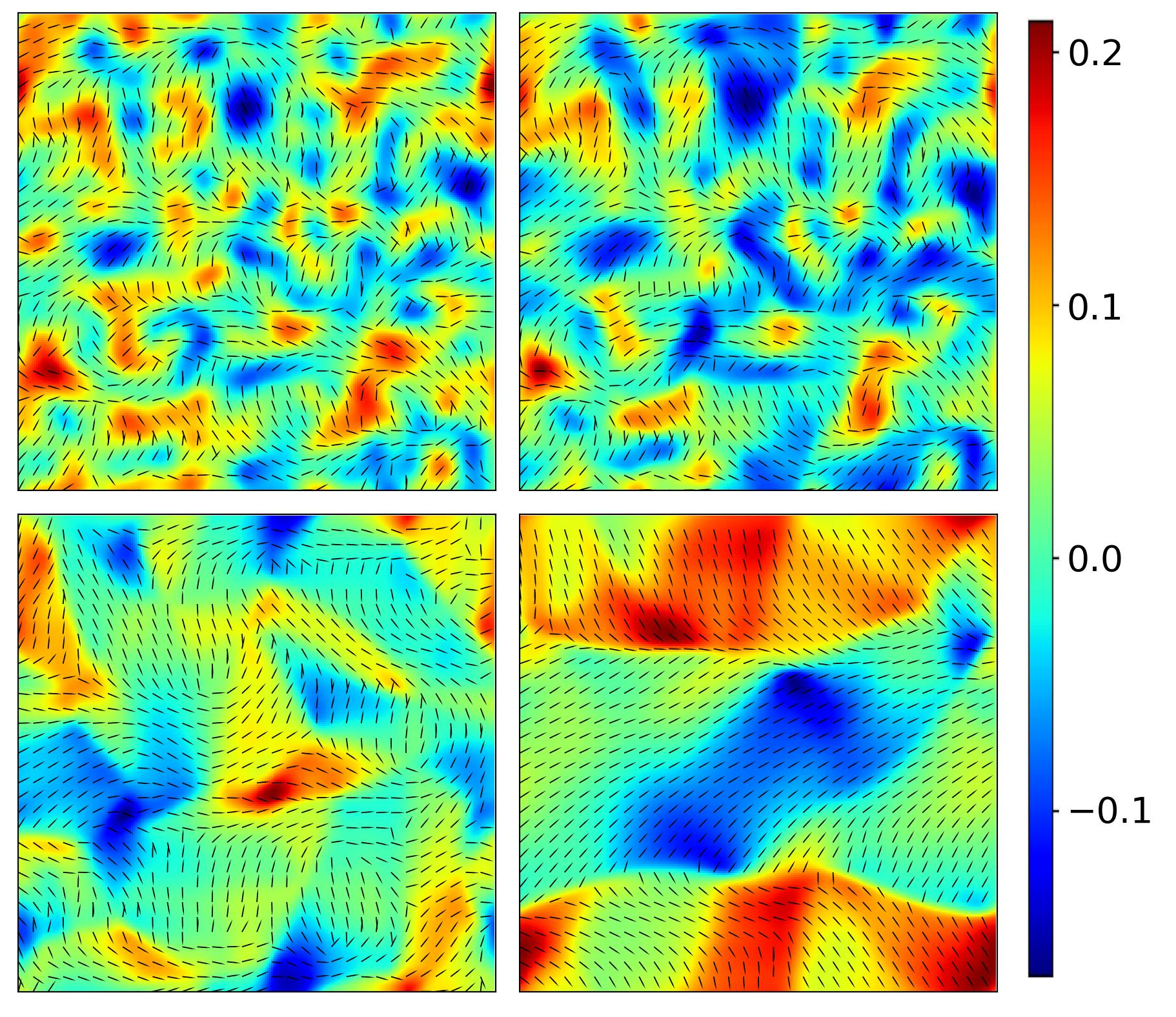}
\caption{\label{fig:coarsen} Coarsening of vortices in time at $L=256$ in absence of friction. The four panels, starting at the top and from left to right, correspond to: (a) $t=7$, (b) $t=10$, (c) $t=40$, and (d) $t=144$. The color indicates vorticity and the lines indicate the orientation of the nematic director.}
\end{figure}

Introducing friction screens the flows and yields a variety of regular emergent structures on scales set by the viscous length $l_\eta$. These include: defect-free bands of opposite flowing material (Fig.~\ref{fig:stat-diag}(f,k)), rectangular lattices of flow vortices with defects arranged in static patterns in the high shear rate regions between vortices of opposite sign (Fig. \ref{fig:stat-diag}(g,h,l,m)), and states of nearly hexagonal vortex lattice that trap pairs of $+1/2$ defects. At the highest activity eventually flows become spatiotemporally chaotic and defects proliferate. The defect separation and the scale of vortical flows are again largely controlled by the viscous screening length, as suggested by experiments~\cite{guillamat2016probing,martinez2021scaling}.


The nearly hexagonal arrangement of vortices shown for instance in Fig.~\ref{fig:stat-diag}(i) \rev{has the interesting property that} all vortices rotate in the same direction, with pairs of $+1/2$ defects trapped in each vortex and rotating to chase each-other tails. It is a state of spontaneously broken symmetry, \rev{as the system will randomly choose either a CW or CCW direction for all defect pairs to rotate around, while surrounded by a continuous space of the opposite vorticity.}
The total vorticity throughout the system \mcm{remains zero}, as it must be for incompressible Stokes' flow. The \mcm{localized} vorticity of the \mcm{``chiral blobs'' trapping defect pairs} is thus compensated by vorticity of opposite sign created in the high shear regions where the isolated $-1/2$ defects are \mcm{located}. Interestingly, the rotation of defect pairs is synchronized, 
giving rise to propagating waves (see video 5). 

States \rev{with a similar chiral structure to the one found here} have been observed before \mcm{deep in the nematic region of chiral} active liquid crystals~\cite{li2020pattern}, \mcm{where chirality is} explicitly \mcm{broken in the} free energy.
\mcm{Here, in contrast}, 
the system  breaks this symmetry spontaneously, choosing a spinning direction.
\mcm{In addition, unlike previous work,}
ordered states \mcm{are found here in the isotropic regime of the passive liquid crystal for intermediate activities, before reaching active turbulence. The range of activity over which ordered structures appear is admittedly narrow, which may have prevented their experimental observation so far.}  

\yzh{Another type of} emergent structures observed in our frictional film resemble a space-extended version of the Ceilidh dancing state \cite{shendruk2017} \mcm{found in active nematics confined to a channel,} where defects march forward while exchanging partners (SI video 4). \mcm{Physical confinement, however,}
requires fine tuning of the channel width such that it can only fit two defects. Here, \mcm{in contrast, the size selection is intrinsic as it is provided by the flow screening length.}

As mentioned in the Introduction, the emergence of coherent flow structures \mcmrev{at finite flow screening length} can be interpreted as a form of phase separation of vorticity, \mcmrev{analogue to MIPS in systems with birth and death~\cite{cates2010}.} 
This is supported by Fig.~\ref{fig:coarsen} that shows how in the absence of substrate frictions regions of 
positive and negative vorticity \mcm{coarsen in time} until \mcmrev{flow consists of two oppositely rotating system-spanning vortices.} 
The dynamics resembles \rev{\mcmrev{qualitatively the} coarsening dynamics of a conserved field} \mcmrev{undergoing bulk phase separation.}
\mcm{Substrate friction plays a role similar to birth/death in MIPS by arresting phase separation~\cite{cates2010}. Friction breaks momentum conservation and selects the size of the flow vortices.  When the flow screening length is comparable to the active length of the unscreened system, the size of emerging structures is cut off though a mechanism that can  be thought as a form of micro phase separation of vorticity.}
We will draw this parallel in a quantitative manner in the following section through linear stability analysis of the dynamics.



\begin{figure}
    \centering
    \includegraphics[scale=0.17]{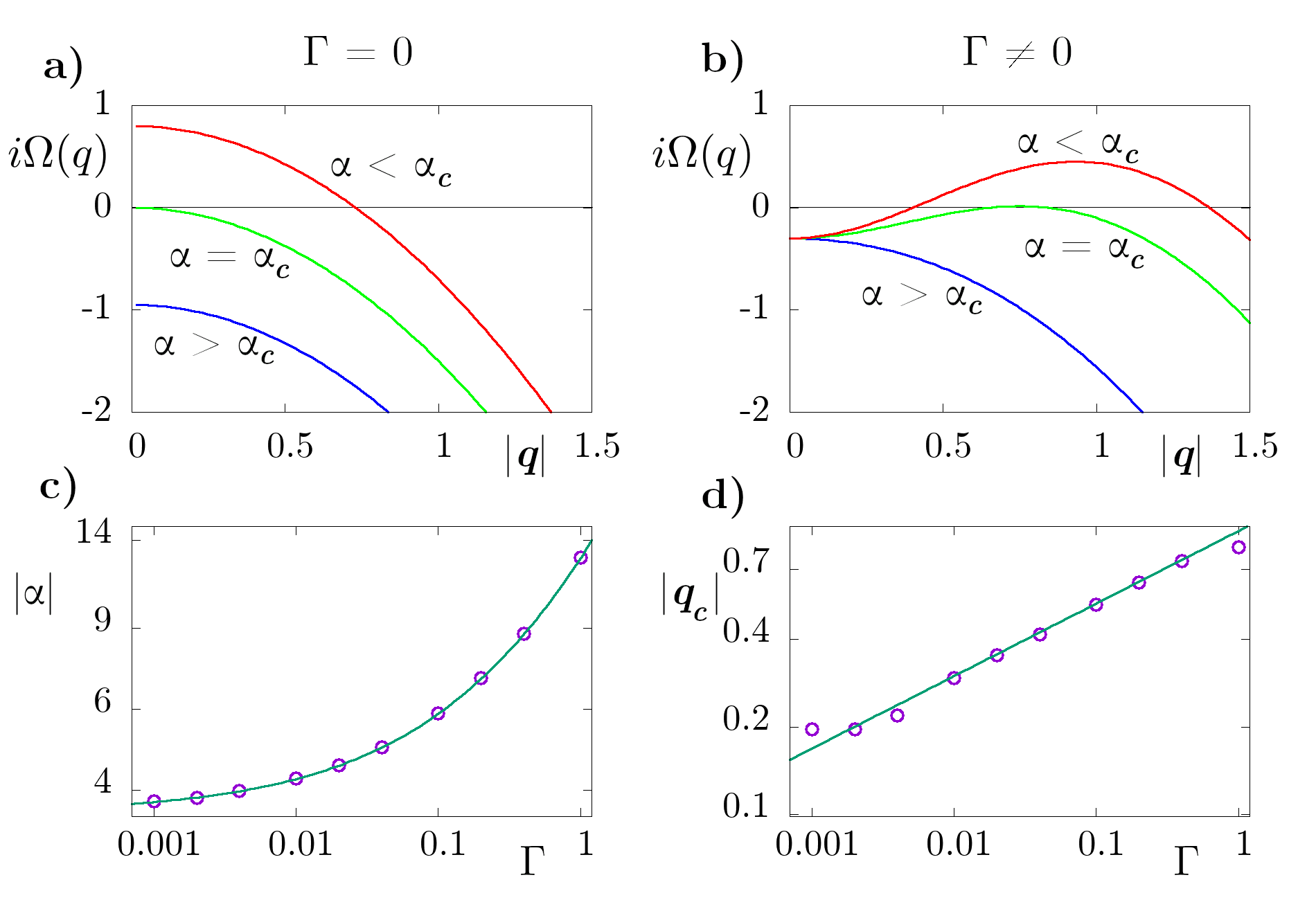}
    \caption{This plot shows sketches of the dispersion relation below and above the critical activity $\alpha_c$ for the cases of no friction (a), and finite friction (b). Plot c) shows the critical activity $|\alpha_c|$ as a function of friction $\Gamma$, while d) shows the most unstable mode as a function of $\Gamma$, where the value of $\alpha$ for each point is the one shown in c) for that friction. The points are results of numerically integrating equations \ref{eq:dtvQ}, while the lines are the predictions of equations \ref{eq:ac} and \ref{eq:qc}.}
    \label{fig:linears}
\end{figure}

\section{Linear Stability analysis}

The phase separation-like dynamics of vorticity can be understood qualitatively by \mcm{examining the  linear} stability of the isotropic state. To do so, we \mcm{linearize Eqs.~\ref{eq:dtQ} and \ref{eq:dtv} about $\mathbf{Q}=\mathbf{v}=0$ and Fourier transform in space. It is useful} to introduce the components of $\Qb$ parallel and perpendicular \mcm{to the wavevector $\mathbf{q}$, given by $\Psi_\parallel = q_iq_j \hat{Q}_{ij}(\mathbf{q})$ and $\Psi_\perp = \epsilon_{ij}q_i q_k \hat{Q}_{jk}(\mathbf{q})$, where $\hat{Q}_{ij}(\mathbf{q})= \int d^dr e^{i \mathbf q\cdot\mathbf{r}}Q_{ij}(r)$} is the Fourier transform of the order parameter. These components decouple and their dynamics is given by
\begin{subequations}
\begin{align}
\label{eq:psipara}
\partial_t\Psi_\parallel &= -\frac{a+Kq^2}{\gamma}\Psi_\parallel\;,\\
\label{eq:psiperp}
\partial_t\Psi_\perp &= -\frac{a+Kq^2}{\gamma}\Psi_\perp-\frac{\lambda q^2}{2}\hat\omega\;,
\end{align}
\end{subequations}
where $\omega=\partial_{x}v_{y}-\partial_{y}v_{x}$ is the flow vorticity. The parallel component $\Psi_\parallel$ 
is always stable, so we ignore it in what follows. To close Eq. \eqref{eq:psiperp}, we consider the Stokes limit for the flow, and drop all nonlinear terms. Taking the curl of Eq. \eqref{eq:dtv} \rev{and transforming to Fourier space gives the vorticity in Fourier space, $\hat\omega$,}
\eq
\label{eq:omega}
\hat\omega=\frac{\alpha+\lambda(a+Kq^2)}{\Gamma+\eta q^2}\Psi_\perp\;.
\eeq
Substituting Eq. \eqref{eq:omega} in Eq. \eqref{eq:psiperp}, we finally arrive at the linearized equation for $\partial_t\Psi_\perp = i\Omega(q)\Psi_\perp$, with
\begin{equation}
\label{eq:psiperp1}
    i\Omega(q) = -\frac{a+K q^2}{\gamma} - \frac{\lambda q^2}{2}\frac{\alpha+\lambda(a+Kq^2)}{\Gamma+\eta q^2}\;.
\end{equation}
\mcm{It is evident} that an extensile stress $\alpha<0$ is the only thing that can render $\partial_t\Psi_\perp$ positive for some wavevectors $q$,  thus 
destabilizing the isotropic state~\cite{srivastava2016,santhosh2020,Adkins2022}. The nature of this activity-driven instability depends \mcm{on the interplay of the two dissipation mechanisms controlled by} substrate friction and viscosity. \mcm{It is instructive to first} analyze the limiting cases $\Gamma\rightarrow 0$ and $\eta\rightarrow 0$. 
When $\Gamma=0$, the dispersion relation of the mode controlling to dynamics of $\Psi_\perp$ is given by
\begin{equation}
\left.i\Omega(q)\right|_{\Gamma=0} = - \left(\frac{\tilde{a}}{\gamma}+\frac{\alpha\lambda}{2\eta}\right)-\frac{\tilde{K}}{\gamma}q^2.
\end{equation}
where $\tilde{a}=a\left(1+\frac{\lambda^2\gamma}{2\eta}\right)$ and $\tilde{K}=K\left(1+\frac{\lambda^2\gamma}{2\eta}\right)$. \rev{Given that} the shear viscosity $\eta$ and nematic rotational viscosity $\gamma$ can be assumed to be of the same order, the dimensionless factor \rev{$\gamma\lambda^2/2\eta$} will just be of order unity. It is evident that in this limit extensile activity can  change the sign of the relaxation rate at $q=0$, effectively driving the system into the nematic state, as evident from the large ordered regions shown in Fig.~\ref{fig:stat-diag}(b).
The dispersion relation is shown in Fig.~\ref{fig:linears}(a). This is a type-III instability according to the classification of Refs.~\cite{cross1993,cross2009}, with system-size spanning emergent structures.  Such long-wave instability is commonly seen in phase separating systems without conserved mass, such as the Allen-Cahn model, and signals coarsening of structures over time, corresponding here to the development of bulk regions of positive and negative vorticity.

Substrate friction changes the nature of this instability, which can be seen in the $\eta=0$ limit of equation \eqref{eq:psiperp1}
\begin{equation}
\left.i\Omega(q)\right|_{\eta=0} = -\frac{a}{\gamma} - \left(\frac{K}{\gamma} + \frac{\alpha\lambda+a\lambda^2}{2\Gamma}\right)q^2 -\frac{K\lambda^2}{2\Gamma}q^4.
\end{equation}
In this case activity renormalizes the stiffness $K$, rendering it negative above a critical value. \mcm{As discussed in detail in Ref.~\cite{srivastava2016}, the system is unstable above the critical activity $\alpha_c^0=\Gamma\left(\sqrt{K/\gamma\lambda^2}+\sqrt{a/2\Gamma}\right)^2$ in a band of wavenumbers.}
Coherent structures emerge at a characteristic length scale corresponding to the most unstable mode \mcm{$q_c^0=l_c^{-1}\left(2\Gamma l_c^2/\gamma\lambda^2\right)^{1/4}$}, in a process \rev{qualitatively} resembling an initial spinodal decomposition, which is then arrested on length scales comparable to $1/q_c^0$ by frictional dissipation that screens the flows. \mcm{The dispersion relation of the modes has the same structure as shown in }
Fig.~\ref{fig:linears}(b) at finite viscosity.

In the presence of both viscosity and substrate friction
the critical activity $\alpha_c$ and most unstable mode $q_c$ at onset can be found from Eq.~ \eqref{eq:psiperp1}, by solving the coupled equations
$\Omega(q_c) = 0$ and $[\partial_q\Omega(q)]_{q=q_c} = 0$, with the result
\begin{equation}\label{eq:ac}
    \alpha_c = \lambda a + \lambda  a\frac{2\eta }{\lambda^2\gamma}\left[1+\frac{l_c^2}{l_\eta^2}+\frac{l_c}{l_\eta}\sqrt{1+\frac{\lambda^2\gamma}{2\eta}}\right]
\end{equation}
and
\begin{equation}\label{eq:qc}
q_c^2 = (l_cl_\eta)^{-1}\left(1+\frac{\lambda^2\gamma}{2\eta}\right)^{-1/2},
\end{equation}
For vanishing friction the instability occurs at
$\alpha_c(\Gamma=0) = a\lambda\left(1 + 2\eta/(\lambda\gamma)\right)$ 
with $q_c=0$~\cite{srivastava2016}.
Substrate friction shifts the instability to higher values of activity and yields a finite length scale $\sim q_c^{-1}$ for emergent structures. 

We have validated the linear theory by measuring the critical activity $\alpha_c$  and the wavenumber $q_c$ of the emerging patterns from simulations. Figures \ref{fig:linears}(c,d) show excellent agreement between simulations and the linear predictions of equations \eqref{eq:ac} and \eqref{eq:qc}. In particular, the linear stability analysis captures the observed dependence of the length scale of the emerging structures with friction, $q_c\sim\Gamma^{1/4}$. Interestingly, increasing activity does not change much the characteristic length scale of the patterns, which is mainly controlled by $l_\eta$, as evident from Fig. \ref{fig:stat-diag}. 


 \section{Vortex lattices}

\rev{Vortex lattices similar to what we have observed in the previous section have been studied before in the context of nonequilibrium hydrodynamics \cite{slomka2017geometry,james2021emergence}. These previous studied offer a path to an analytical description of these lattices that we offer in this section.}

The emergence of lattice structures of flow vortices can also be rationalized in terms of linear hydrodynamics. \mcm{The combination of Eqs.~ \eqref{eq:psiperp} and \eqref{eq:omega} implies that the relaxation rate of vorticity fluctuations is determined by the same dispersion relation that controls  $\Psi_\perp$, given by Eq. \eqref{eq:psiperp1}.}
Expanding the mode for $q\ll l_\eta^{-1}$
gives a simple linear equation commonly seen in models of pattern-formation, $\partial_t\hat\omega = i\Omega_4(q)\hat\omega$, with
\begin{equation}
    \label{eq:omega_q4}
    i\Omega_4(q) =  -\left[\tau_c^{-1} + k_2q^2 + k_4q^4\right] + O(q^6),
\end{equation}
where $k_2=K/\gamma + \lambda(\alpha+a\lambda)/(2\Gamma)$ and $k_4=(-\alpha\eta\lambda + K\Gamma\lambda^2-a\eta\lambda^2)/(2\Gamma^2)$ both depend on activity. Such an equation has been used as a minimal model to study emergent structures and dynamics in active fluids, where
 it has been shown to
 support various time-independent solutions in the form of vortex lattices~\cite{slomka2017geometry} when $k_2<0$. This explains the origin of the periodic structures observed in our simulations. Specifically, an extensile active stress combined with flow alignment can give rise to a negative effective stiffness~\cite{srivastava2016} - effectively a negative effective viscosity when applied to flow.
 Since modes are always damped at very small and large $q$, at the onset of instability only a small interval of wavenumbers become unstable (Fig.~\ref{fig:linears}(b)) near the wavenumber of the most unstable mode $q_c$ (Eq.~\eqref{eq:qc}), which therefore controls the scale of the periodic structures observed.

\begin{figure}[t]
    \centering
    \hspace*{-0.2cm}\includegraphics[scale=0.43]{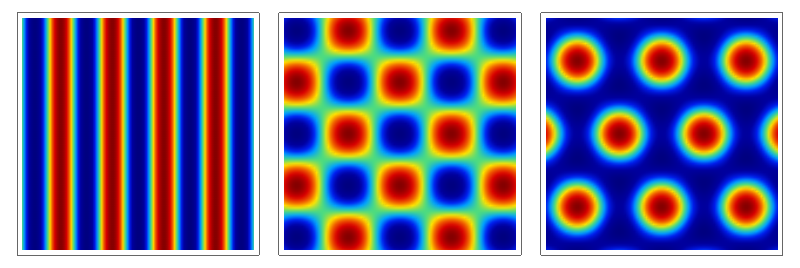}
    \caption{Different states that can be built as static vorticity structures from solutions of Eq.~\ref{eq:lattices}, corresponding to those  observed numerically, for both finite $\Gamma$ and $a$ (see Fig \ref{fig:stat-diag}). The amplitude in this plot has been chosen to be $1$, as it is arbitrary when built from equation \eqref{eq:lattices}.}
    \label{fig:lattices}
\end{figure}

To find the static solutions that correspond to vorticity lattices, we use the following ansatz for the stream function $\psi$, defined by $\omega = -\nabla^2\psi$, in polar coordinates \cite{slomka2017geometry}
\begin{equation}
    \label{eq:lattices}
    \psi(r,\theta) = \int d\phi \hat\psi(\phi) e^{iq r \cos(\theta-\phi)}.
\end{equation}
Different lattices can be constructed by choosing $\hat\psi(\phi)$ to be the sum of different modes in the unit circle. For instance, we can form the band configuration by choosing two symmetric modes, i.e. $\hat\psi = \delta(\phi) - \delta(\phi-\pi)$, which, in Cartesian coordinates, gives a band solution $\psi(x,y) = \cos(q x)$. To build a square lattice, we choose four modes along the unit circle $\hat\psi(\phi) = \delta(\phi-\pi/4)+\delta(\phi-3\pi/4)+\delta(\phi-5\pi/4)+\delta(\phi-7\pi/4)$, giving the stream function for the square lattice $\psi_4(x,y) = \cos(k x)\cos(k y)$. Similarly, if we choose six equidistant points, we obtain the stream function for the hexagonal lattice $\psi_6(x,y) = \cos \left(\sqrt{3} x/2\right) \cos \left(y/2\right)-\cos (y)/2$. These three main configurations are plotted in Fig.~\ref{fig:lattices}, corresponding  to the ones observed numerically for low activity in figure \ref{fig:stat-diag}.

\begin{figure}[t]
    \centering
    \hspace*{-0.2cm}\includegraphics[scale=0.4]{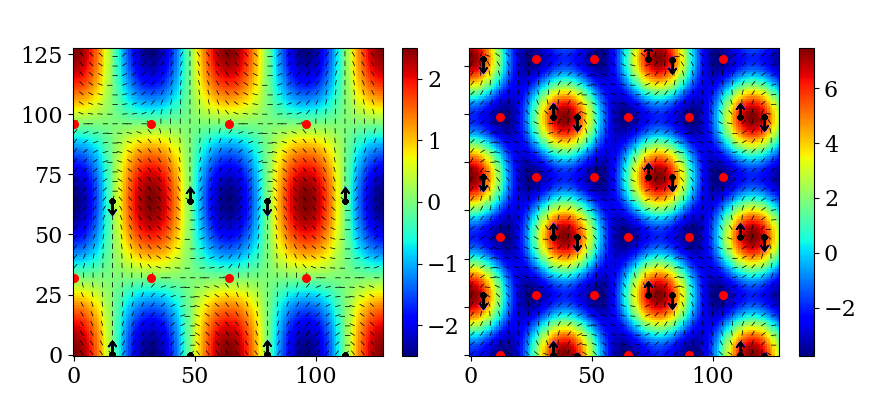}
    \caption{Nematic texture generated by the ansatz $\psi_4$ and $\psi_6$ for the streamfunction described in the main text. The color represents the vorticity. 
    Here the dots denote the $-1/2$ defects, while the arrows represent the polarization of the $+1/2$ defects. The parameter $l$ has been chosen as $l=2$ and $l=1.1$ for the rectangular and hexagonal lattices respectively.}
    \label{fig:solution_lattice}
\end{figure}

When inserted in Eq.~\eqref{eq:dtQ}, and assuming \rev{a static state, in which $D_t Q_{ij}=0$}, the stream functions generated by this method also reproduce the liquid crystalline textures observed numerically and shown in in Fig.~\ref{fig:stat-diag}. 
The solution is trivial in the case of bands, as has been found before \cite{Vafa2021}, with the director oriented $45$ degrees with respect to the bands, and rotated $90$ degrees from band to band.
For the structures shown for instance in Fig.~\ref{fig:stat-diag}(g,h), the defects are organized in a rectangular, rather than square lattice with $D_4$ symmetry.
A perfectly square lattice of vorticity will not therefore reproduce the observed liquid crystal texture. 
We adjust the stream function corresponding to a square lattice by using two different wavevectors along orthogonal coordinate directions, i.e., $\psi_4 = \cos(k_1x)\cos(l k_1 y)$, where the parameter $l\neq1$ describes the ratio between the two axis' wavelengths. The velocity calculated from this stream function as $v_i = \epsilon_{ij}\partial_j\psi$ is inserted in Eq.~$\eqref{eq:dtQ}$, which is then solved numerically  with $D_t Q_{ij}=0$. This gives  the nematic texture shown in Fig.~\ref{fig:solution_lattice}a), with $+1/2$ defects aligned across the direction of shorter wavelength of vorticity, and $-1/2$ defects across the direction of longer wavelength, in agreement with what obtained from simulations.

Likewise, to obtain the hexagonal lattice structure shown in Fig.~\ref{fig:stat-diag}(i,n) we use a stream function with hexagonal and two different wavevectors, $\psi_6 = \cos \left(k_1\sqrt{3} x/2\right) \cos \left(l k_1 y/2\right)-\cos (l k_1 y)/2$. 
Inserting  the velocity field generated by this stream function into Eq.~\eqref{eq:dtQ} and solving numerically yields the liquid crystal texture observed in simulations, with pairs of $+1/2$ defects trapped within each vortex, and  $-1/2$ defects trapped at the stagnation points where  opposite vorticities meet, forming a hexagonal lattice around each flow vortex (see Fig.~\ref{fig:solution_lattice}b). 
The fact that we can reproduce the nematic texture from a simple ansatz for the vortical flows demonstrates the direct connection between flow and texture that allows us to tune flows by controlling defects.

Finally, we stress that although the rectangular lattice is a static solution to equations \eqref{eq:dtvQ}, the hexagonal lattice is not. In the hexagonal case, the $+1/2$ are motile and continuously spin chasing each other tail. The stream function would then be a time-dependent version of $\psi_6$, in which the wavevectors  make the lattice continuously rotate as  seen in video 5. We leave the full characterization of these dynamical states for future work.

\section{Discussion and Conclusion}

The capability of producing diverse forms of collective motion is a distinct feature of active matter. 
\mcm{Here we have demonstrated that activity and substrate friction can be used as handles for generating highly organized and controllable structures in the isotropic phase of active liquid crystals.  
These coherent structures exist in an intermediate range of activity, out of which the system becomes either isotropic quiescent or chaotic and require that the nematic coherence length $l_c$ that controls spatial variations of the order parameter be smaller than or at most comparable to the screening length of the flow $l_\eta$.  
The range of activity where we observe coherent structure is wider when the system is close to the passive critical point. This can be achieved, while maintaining $l_c<l_\eta$ by either increasing the nematic stiffness $K$ or the nonlinear damping $b$. 
Tuning} substrate friction allows us to change the \mcm{characteristic scale of the flow and associated texture} continuously, as well as the symmetry of the  vortex lattice, with self-organized structures that resemble those found in  phenomenological models of active pattern formation~\cite{oza2016antipolar}.
\mcm{This tunability may provide a path for the application of active fluids to microfluidics} \mcmrev{for instance by tuning friction as realized in Ref. \cite{thijssen2021submersed}}.

\mcm{Perhaps the most intriguing structure is the lattice of rotating chiral blobs that trap pairs of topological defects. Each of this chiral units exhibit a structure that resembles that of nematic liquid crystals confined to circular wells. In our case confinement is effectively provided by the screening length $l_\eta$, hence is emergent. }

\mcm{The behavior observed here can be interpreted in analogy with MIPS as a phase separation of vortical flow. In the absence of friction, the system organizes in system-spanning counter-rotating vortices through a dynamics that resembles spinodal decomposition. 
Substrate friction screens the flow and breaks momentum conservation, arresting this vortical phase separation and organizing the system in ordered flow structures. Friction therefore plays a role very similar to that of the breaking of mass conservation in MIPS~\cite{cates2010}. }

\mcm{It would be interesting to examine} the coarsening of vortices over time to extract scaling laws analogue to those that hold in equilibrium phase separation, as well as study the dynamics and stability of the defect/vortex lattices. \mcm{Another direction left for future studies is the role of confinement provided either by physical boundaries or deformable interfaces.} 
The interplay between activity, flow screening and boundaries could lead to even richer behaviors that could be exploited to create smart functional materials.

\section*{Conflicts of interest}
There are no conflicts of interest.

\mcm{\acknowledgements{This work was supported by the National Science Foundation award No. DMR-2041459.}}

\bibliography{Vorticity_PS_clean_copy}

\providecommand*{\mcitethebibliography}{\thebibliography}
\csname @ifundefined\endcsname{endmcitethebibliography}
{\let\endmcitethebibliography\endthebibliography}{}
\begin{mcitethebibliography}{44}
\providecommand*{\natexlab}[1]{#1}
\providecommand*{\mciteSetBstSublistMode}[1]{}
\providecommand*{\mciteSetBstMaxWidthForm}[2]{}
\providecommand*{\mciteBstWouldAddEndPuncttrue}
  {\def\EndOfBibitem{\unskip.}}
\providecommand*{\mciteBstWouldAddEndPunctfalse}
  {\let\EndOfBibitem\relax}
\providecommand*{\mciteSetBstMidEndSepPunct}[3]{}
\providecommand*{\mciteSetBstSublistLabelBeginEnd}[3]{}
\providecommand*{\EndOfBibitem}{}
\mciteSetBstSublistMode{f}
\mciteSetBstMaxWidthForm{subitem}
{(\emph{\alph{mcitesubitemcount}})}
\mciteSetBstSublistLabelBeginEnd{\mcitemaxwidthsubitemform\space}
{\relax}{\relax}

\bibitem[Aditi~Simha and Ramaswamy(2002)]{aditi2002}
R.~Aditi~Simha and S.~Ramaswamy, \emph{Phys. Rev. Lett.}, 2002, \textbf{89},
  058101\relax
\mciteBstWouldAddEndPuncttrue
\mciteSetBstMidEndSepPunct{\mcitedefaultmidpunct}
{\mcitedefaultendpunct}{\mcitedefaultseppunct}\relax
\EndOfBibitem
\bibitem[Marchetti \emph{et~al.}(2012)Marchetti, Joanny, Ramaswamy, Liverpool,
  Prost, Rao, and Simha]{marchetti2012soft}
M.~Marchetti, J.-F. Joanny, S.~Ramaswamy, T.~Liverpool, J.~Prost, M.~Rao and
  R.~A. Simha, \emph{arXiv preprint arXiv:1207.2929}, 2012\relax
\mciteBstWouldAddEndPuncttrue
\mciteSetBstMidEndSepPunct{\mcitedefaultmidpunct}
{\mcitedefaultendpunct}{\mcitedefaultseppunct}\relax
\EndOfBibitem
\bibitem[Alert \emph{et~al.}(2022)Alert, Casademunt, and
  Joanny]{alert2022active}
R.~Alert, J.~Casademunt and J.-F. Joanny, \emph{Annual Review of Condensed
  Matter Physics}, 2022, \textbf{13}, 143--170\relax
\mciteBstWouldAddEndPuncttrue
\mciteSetBstMidEndSepPunct{\mcitedefaultmidpunct}
{\mcitedefaultendpunct}{\mcitedefaultseppunct}\relax
\EndOfBibitem
\bibitem[Lemma \emph{et~al.}(2021)Lemma, Norton, Tayar, DeCamp, Aghvami,
  Fraden, Hagan, and Dogic]{lemma2021multiscale}
L.~M. Lemma, M.~M. Norton, A.~M. Tayar, S.~J. DeCamp, S.~A. Aghvami, S.~Fraden,
  M.~F. Hagan and Z.~Dogic, \emph{Physical Review Letters}, 2021, \textbf{127},
  148001\relax
\mciteBstWouldAddEndPuncttrue
\mciteSetBstMidEndSepPunct{\mcitedefaultmidpunct}
{\mcitedefaultendpunct}{\mcitedefaultseppunct}\relax
\EndOfBibitem
\bibitem[Tayar \emph{et~al.}(2021)Tayar, Hagan, and Dogic]{tayar2021active}
A.~M. Tayar, M.~F. Hagan and Z.~Dogic, \emph{Proceedings of the National
  Academy of Sciences}, 2021, \textbf{118}, e2102873118\relax
\mciteBstWouldAddEndPuncttrue
\mciteSetBstMidEndSepPunct{\mcitedefaultmidpunct}
{\mcitedefaultendpunct}{\mcitedefaultseppunct}\relax
\EndOfBibitem
\bibitem[Tayar \emph{et~al.}(2022)Tayar, Caballaro, Anderberg, Saleh,
  Marchetti, and Dogic]{tayar2022controlling}
A.~M. Tayar, F.~Caballaro, T.~Anderberg, O.~A. Saleh, M.~C. Marchetti and
  Z.~Dogic, \emph{arXiv preprint arXiv:2208.12769}, 2022\relax
\mciteBstWouldAddEndPuncttrue
\mciteSetBstMidEndSepPunct{\mcitedefaultmidpunct}
{\mcitedefaultendpunct}{\mcitedefaultseppunct}\relax
\EndOfBibitem
\bibitem[Sanchez \emph{et~al.}(2011)Sanchez, Welch, Nicastro, and
  Dogic]{sanchez2011cilia}
T.~Sanchez, D.~Welch, D.~Nicastro and Z.~Dogic, \emph{Science}, 2011,
  \textbf{333}, 456--459\relax
\mciteBstWouldAddEndPuncttrue
\mciteSetBstMidEndSepPunct{\mcitedefaultmidpunct}
{\mcitedefaultendpunct}{\mcitedefaultseppunct}\relax
\EndOfBibitem
\bibitem[Adkins \emph{et~al.}(2022)Adkins, Kolvin, You, Witthaus, Marchetti,
  and Dogic]{Adkins2022}
R.~Adkins, I.~Kolvin, Z.~Y. You, S.~Witthaus, M.~C. Marchetti and Z.~Dogic,
  \emph{Science}, 2022\relax
\mciteBstWouldAddEndPuncttrue
\mciteSetBstMidEndSepPunct{\mcitedefaultmidpunct}
{\mcitedefaultendpunct}{\mcitedefaultseppunct}\relax
\EndOfBibitem
\bibitem[Giomi \emph{et~al.}(2011)Giomi, Mahadevan, Chakraborty, and
  Hagan]{giomi2011}
L.~Giomi, L.~Mahadevan, B.~Chakraborty and M.~F. Hagan, \emph{Phys. Rev.
  Lett.}, 2011, \textbf{106}, 218101\relax
\mciteBstWouldAddEndPuncttrue
\mciteSetBstMidEndSepPunct{\mcitedefaultmidpunct}
{\mcitedefaultendpunct}{\mcitedefaultseppunct}\relax
\EndOfBibitem
\bibitem[Giomi \emph{et~al.}(2012)Giomi, Mahadevan, Chakraborty, and
  Hagan]{giomi2012}
L.~Giomi, L.~Mahadevan, B.~Chakraborty and M.~Hagan, \emph{Nonlinearity}, 2012,
  \textbf{25}, 2245\relax
\mciteBstWouldAddEndPuncttrue
\mciteSetBstMidEndSepPunct{\mcitedefaultmidpunct}
{\mcitedefaultendpunct}{\mcitedefaultseppunct}\relax
\EndOfBibitem
\bibitem[Doostmohammadi \emph{et~al.}(2017)Doostmohammadi, Shendruk, Thijssen,
  and Yeomans]{doostmohammadi2017}
A.~Doostmohammadi, T.~N. Shendruk, K.~Thijssen and J.~M. Yeomans, \emph{Nature
  communications}, 2017, \textbf{8}, 1--7\relax
\mciteBstWouldAddEndPuncttrue
\mciteSetBstMidEndSepPunct{\mcitedefaultmidpunct}
{\mcitedefaultendpunct}{\mcitedefaultseppunct}\relax
\EndOfBibitem
\bibitem[Shendruk \emph{et~al.}(2017)Shendruk, Doostmohammadi, Thijssen, and
  Yeomans]{shendruk2017}
T.~N. Shendruk, A.~Doostmohammadi, K.~Thijssen and J.~M. Yeomans, \emph{Soft
  Matter}, 2017, \textbf{13}, 3853--3862\relax
\mciteBstWouldAddEndPuncttrue
\mciteSetBstMidEndSepPunct{\mcitedefaultmidpunct}
{\mcitedefaultendpunct}{\mcitedefaultseppunct}\relax
\EndOfBibitem
\bibitem[Opathalage \emph{et~al.}(2019)Opathalage, Norton, Juniper, Langeslay,
  Aghvami, Fraden, and Dogic]{opathalage2019}
A.~Opathalage, M.~M. Norton, M.~P. Juniper, B.~Langeslay, S.~A. Aghvami,
  S.~Fraden and Z.~Dogic, \emph{Proceedings of the National Academy of
  Sciences}, 2019, \textbf{116}, 4788--4797\relax
\mciteBstWouldAddEndPuncttrue
\mciteSetBstMidEndSepPunct{\mcitedefaultmidpunct}
{\mcitedefaultendpunct}{\mcitedefaultseppunct}\relax
\EndOfBibitem
\bibitem[Hardo{\"u}in \emph{et~al.}(2019)Hardo{\"u}in, Hughes, Doostmohammadi,
  Laurent, Lopez-Leon, Yeomans, Ign{\'e}s-Mullol, and Sagu{\'e}s]{hardouin2019}
J.~Hardo{\"u}in, R.~Hughes, A.~Doostmohammadi, J.~Laurent, T.~Lopez-Leon, J.~M.
  Yeomans, J.~Ign{\'e}s-Mullol and F.~Sagu{\'e}s, \emph{Communications
  Physics}, 2019, \textbf{2}, 1--9\relax
\mciteBstWouldAddEndPuncttrue
\mciteSetBstMidEndSepPunct{\mcitedefaultmidpunct}
{\mcitedefaultendpunct}{\mcitedefaultseppunct}\relax
\EndOfBibitem
\bibitem[Kempf \emph{et~al.}(2019)Kempf, Mueller, Frey, Yeomans, and
  Doostmohammadi]{kempf2019active}
F.~Kempf, R.~Mueller, E.~Frey, J.~M. Yeomans and A.~Doostmohammadi, \emph{Soft
  matter}, 2019, \textbf{15}, 7538--7546\relax
\mciteBstWouldAddEndPuncttrue
\mciteSetBstMidEndSepPunct{\mcitedefaultmidpunct}
{\mcitedefaultendpunct}{\mcitedefaultseppunct}\relax
\EndOfBibitem
\bibitem[Caballero and Marchetti(2022)]{caballero2022activity}
F.~Caballero and M.~C. Marchetti, \emph{Physical Review Letters}, 2022,
  \textbf{129}, 268002\relax
\mciteBstWouldAddEndPuncttrue
\mciteSetBstMidEndSepPunct{\mcitedefaultmidpunct}
{\mcitedefaultendpunct}{\mcitedefaultseppunct}\relax
\EndOfBibitem
\bibitem[Bowick \emph{et~al.}(2022)Bowick, Fakhri, Marchetti, and
  Ramaswamy]{bowick2022symmetry}
M.~J. Bowick, N.~Fakhri, M.~C. Marchetti and S.~Ramaswamy, \emph{Physical
  Review X}, 2022, \textbf{12}, 010501\relax
\mciteBstWouldAddEndPuncttrue
\mciteSetBstMidEndSepPunct{\mcitedefaultmidpunct}
{\mcitedefaultendpunct}{\mcitedefaultseppunct}\relax
\EndOfBibitem
\bibitem[Doostmohammadi \emph{et~al.}(2016)Doostmohammadi, Adamer, Thampi, and
  Yeomans]{doostmohammadi2016}
A.~Doostmohammadi, M.~F. Adamer, S.~P. Thampi and J.~M. Yeomans, \emph{Nature
  communications}, 2016, \textbf{7}, 1--9\relax
\mciteBstWouldAddEndPuncttrue
\mciteSetBstMidEndSepPunct{\mcitedefaultmidpunct}
{\mcitedefaultendpunct}{\mcitedefaultseppunct}\relax
\EndOfBibitem
\bibitem[Guillamat \emph{et~al.}(2016)Guillamat, Ign{\'e}s-Mullol, and
  Sagu{\'e}s]{guillamat2016}
P.~Guillamat, J.~Ign{\'e}s-Mullol and F.~Sagu{\'e}s, \emph{Proceedings of the
  National Academy of Sciences}, 2016, \textbf{113}, 5498--5502\relax
\mciteBstWouldAddEndPuncttrue
\mciteSetBstMidEndSepPunct{\mcitedefaultmidpunct}
{\mcitedefaultendpunct}{\mcitedefaultseppunct}\relax
\EndOfBibitem
\bibitem[Guillamat \emph{et~al.}(2017)Guillamat, Ign{\'e}s-Mullol, and
  Sagu{\'e}s]{guillamat2017}
P.~Guillamat, J.~Ign{\'e}s-Mullol and F.~Sagu{\'e}s, \emph{Nature
  communications}, 2017, \textbf{8}, 1--8\relax
\mciteBstWouldAddEndPuncttrue
\mciteSetBstMidEndSepPunct{\mcitedefaultmidpunct}
{\mcitedefaultendpunct}{\mcitedefaultseppunct}\relax
\EndOfBibitem
\bibitem[Thijssen \emph{et~al.}(2020)Thijssen, Metselaar, Yeomans, and
  Doostmohammadi]{thijssen2020}
K.~Thijssen, L.~Metselaar, J.~M. Yeomans and A.~Doostmohammadi, \emph{Soft
  Matter}, 2020, \textbf{16}, 2065--2074\relax
\mciteBstWouldAddEndPuncttrue
\mciteSetBstMidEndSepPunct{\mcitedefaultmidpunct}
{\mcitedefaultendpunct}{\mcitedefaultseppunct}\relax
\EndOfBibitem
\bibitem[Mart{\'\i}nez-Prat \emph{et~al.}(2021)Mart{\'\i}nez-Prat, Alert, Meng,
  Ign{\'e}s-Mullol, Joanny, Casademunt, Golestanian, and
  Sagu{\'e}s]{martinez2021scaling}
B.~Mart{\'\i}nez-Prat, R.~Alert, F.~Meng, J.~Ign{\'e}s-Mullol, J.-F. Joanny,
  J.~Casademunt, R.~Golestanian and F.~Sagu{\'e}s, \emph{Physical Review X},
  2021, \textbf{11}, 031065\relax
\mciteBstWouldAddEndPuncttrue
\mciteSetBstMidEndSepPunct{\mcitedefaultmidpunct}
{\mcitedefaultendpunct}{\mcitedefaultseppunct}\relax
\EndOfBibitem
\bibitem[Thijssen \emph{et~al.}(2020)Thijssen, Nejad, and
  Yeomans]{thijssen2020role}
K.~Thijssen, M.~R. Nejad and J.~M. Yeomans, \emph{Physical Review Letters},
  2020, \textbf{125}, 218004\relax
\mciteBstWouldAddEndPuncttrue
\mciteSetBstMidEndSepPunct{\mcitedefaultmidpunct}
{\mcitedefaultendpunct}{\mcitedefaultseppunct}\relax
\EndOfBibitem
\bibitem[Thampi \emph{et~al.}(2014)Thampi, Golestanian, and
  Yeomans]{thampi2014active}
S.~P. Thampi, R.~Golestanian and J.~M. Yeomans, \emph{Physical Review E}, 2014,
  \textbf{90}, 062307\relax
\mciteBstWouldAddEndPuncttrue
\mciteSetBstMidEndSepPunct{\mcitedefaultmidpunct}
{\mcitedefaultendpunct}{\mcitedefaultseppunct}\relax
\EndOfBibitem
\bibitem[Chandragiri \emph{et~al.}(2020)Chandragiri, Doostmohammadi, Yeomans,
  and Thampi]{chandragiri2020flow}
S.~Chandragiri, A.~Doostmohammadi, J.~M. Yeomans and S.~P. Thampi,
  \emph{Physical Review Letters}, 2020, \textbf{125}, 148002\relax
\mciteBstWouldAddEndPuncttrue
\mciteSetBstMidEndSepPunct{\mcitedefaultmidpunct}
{\mcitedefaultendpunct}{\mcitedefaultseppunct}\relax
\EndOfBibitem
\bibitem[Keogh \emph{et~al.}(2022)Keogh, Chandragiri, Loewe, Ala-Nissila,
  Thampi, and Shendruk]{keogh2022helical}
R.~R. Keogh, S.~Chandragiri, B.~Loewe, T.~Ala-Nissila, S.~P. Thampi and T.~N.
  Shendruk, \emph{Physical Review E}, 2022, \textbf{106}, L012602\relax
\mciteBstWouldAddEndPuncttrue
\mciteSetBstMidEndSepPunct{\mcitedefaultmidpunct}
{\mcitedefaultendpunct}{\mcitedefaultseppunct}\relax
\EndOfBibitem
\bibitem[Thijssen \emph{et~al.}(2021)Thijssen, Khaladj, Aghvami, Gharbi,
  Fraden, Yeomans, Hirst, and Shendruk]{thijssen2021submersed}
K.~Thijssen, D.~A. Khaladj, S.~A. Aghvami, M.~A. Gharbi, S.~Fraden, J.~M.
  Yeomans, L.~S. Hirst and T.~N. Shendruk, \emph{Proceedings of the National
  Academy of Sciences}, 2021, \textbf{118}, e2106038118\relax
\mciteBstWouldAddEndPuncttrue
\mciteSetBstMidEndSepPunct{\mcitedefaultmidpunct}
{\mcitedefaultendpunct}{\mcitedefaultseppunct}\relax
\EndOfBibitem
\bibitem[Thijssen \emph{et~al.}(2020)Thijssen, Metselaar, Yeomans, and
  Doostmohammadi]{thijssen2020active}
K.~Thijssen, L.~Metselaar, J.~M. Yeomans and A.~Doostmohammadi, \emph{Soft
  Matter}, 2020, \textbf{16}, 2065--2074\relax
\mciteBstWouldAddEndPuncttrue
\mciteSetBstMidEndSepPunct{\mcitedefaultmidpunct}
{\mcitedefaultendpunct}{\mcitedefaultseppunct}\relax
\EndOfBibitem
\bibitem[Cates \emph{et~al.}(2010)Cates, Marenduzzo, Pagonabarraga, and
  Tailleur]{cates2010}
M.~E. Cates, D.~Marenduzzo, I.~Pagonabarraga and J.~Tailleur, \emph{Proceedings
  of the National Academy of Sciences}, 2010, \textbf{107}, 11715--11720\relax
\mciteBstWouldAddEndPuncttrue
\mciteSetBstMidEndSepPunct{\mcitedefaultmidpunct}
{\mcitedefaultendpunct}{\mcitedefaultseppunct}\relax
\EndOfBibitem
\bibitem[Tailleur and Cates(2008)]{tailleur2008statistical}
J.~Tailleur and M.~Cates, \emph{Physical review letters}, 2008, \textbf{100},
  218103\relax
\mciteBstWouldAddEndPuncttrue
\mciteSetBstMidEndSepPunct{\mcitedefaultmidpunct}
{\mcitedefaultendpunct}{\mcitedefaultseppunct}\relax
\EndOfBibitem
\bibitem[Fily and Marchetti(2012)]{fily2012athermal}
Y.~Fily and M.~C. Marchetti, \emph{Physical review letters}, 2012,
  \textbf{108}, 235702\relax
\mciteBstWouldAddEndPuncttrue
\mciteSetBstMidEndSepPunct{\mcitedefaultmidpunct}
{\mcitedefaultendpunct}{\mcitedefaultseppunct}\relax
\EndOfBibitem
\bibitem[Redner \emph{et~al.}(2013)Redner, Hagan, and
  Baskaran]{redner2013structure}
G.~S. Redner, M.~F. Hagan and A.~Baskaran, \emph{Physical review letters},
  2013, \textbf{110}, 055701\relax
\mciteBstWouldAddEndPuncttrue
\mciteSetBstMidEndSepPunct{\mcitedefaultmidpunct}
{\mcitedefaultendpunct}{\mcitedefaultseppunct}\relax
\EndOfBibitem
\bibitem[Cates and Tailleur(2015)]{cates2015motility}
M.~E. Cates and J.~Tailleur, \emph{Annu. Rev. Condens. Matter Phys.}, 2015,
  \textbf{6}, 219--244\relax
\mciteBstWouldAddEndPuncttrue
\mciteSetBstMidEndSepPunct{\mcitedefaultmidpunct}
{\mcitedefaultendpunct}{\mcitedefaultseppunct}\relax
\EndOfBibitem
\bibitem[S{\l}omka and Dunkel(2017)]{slomka2017geometry}
J.~S{\l}omka and J.~Dunkel, \emph{Physical Review Fluids}, 2017, \textbf{2},
  043102\relax
\mciteBstWouldAddEndPuncttrue
\mciteSetBstMidEndSepPunct{\mcitedefaultmidpunct}
{\mcitedefaultendpunct}{\mcitedefaultseppunct}\relax
\EndOfBibitem
\bibitem[Srivastava \emph{et~al.}(2016)Srivastava, Mishra, and
  Marchetti]{srivastava2016}
P.~Srivastava, P.~Mishra and M.~C. Marchetti, \emph{Soft Matter}, 2016,
  \textbf{12}, 8214--8225\relax
\mciteBstWouldAddEndPuncttrue
\mciteSetBstMidEndSepPunct{\mcitedefaultmidpunct}
{\mcitedefaultendpunct}{\mcitedefaultseppunct}\relax
\EndOfBibitem
\bibitem[Putzig \emph{et~al.}(2016)Putzig, Redner, Baskaran, and
  Baskaran]{putzig2016instabilities}
E.~Putzig, G.~S. Redner, A.~Baskaran and A.~Baskaran, \emph{Soft matter}, 2016,
  \textbf{12}, 3854--3859\relax
\mciteBstWouldAddEndPuncttrue
\mciteSetBstMidEndSepPunct{\mcitedefaultmidpunct}
{\mcitedefaultendpunct}{\mcitedefaultseppunct}\relax
\EndOfBibitem
\bibitem[Oza and Dunkel(2016)]{oza2016antipolar}
A.~U. Oza and J.~Dunkel, \emph{New Journal of Physics}, 2016, \textbf{18},
  093006\relax
\mciteBstWouldAddEndPuncttrue
\mciteSetBstMidEndSepPunct{\mcitedefaultmidpunct}
{\mcitedefaultendpunct}{\mcitedefaultseppunct}\relax
\EndOfBibitem
\bibitem[Santhosh \emph{et~al.}(2020)Santhosh, Nejad, Doostmohammadi, Yeomans,
  and Thampi]{santhosh2020}
S.~Santhosh, M.~R. Nejad, A.~Doostmohammadi, J.~M. Yeomans and S.~P. Thampi,
  \emph{Journal of Statistical Physics}, 2020, \textbf{180}, 699–709\relax
\mciteBstWouldAddEndPuncttrue
\mciteSetBstMidEndSepPunct{\mcitedefaultmidpunct}
{\mcitedefaultendpunct}{\mcitedefaultseppunct}\relax
\EndOfBibitem
\bibitem[Vafa \emph{et~al.}(2021)Vafa, Bowick, Shraiman, and
  Marchetti]{Vafa2021}
F.~Vafa, M.~J. Bowick, B.~I. Shraiman and M.~C. Marchetti, \emph{Soft Matter},
  2021, \textbf{17}, 3068--3073\relax
\mciteBstWouldAddEndPuncttrue
\mciteSetBstMidEndSepPunct{\mcitedefaultmidpunct}
{\mcitedefaultendpunct}{\mcitedefaultseppunct}\relax
\EndOfBibitem
\bibitem[Guillamat \emph{et~al.}(2016)Guillamat, Ign{\'e}s-Mullol, Shankar,
  Marchetti, and Sagu{\'e}s]{guillamat2016probing}
P.~Guillamat, J.~Ign{\'e}s-Mullol, S.~Shankar, M.~C. Marchetti and
  F.~Sagu{\'e}s, \emph{Physical review E}, 2016, \textbf{94}, 060602\relax
\mciteBstWouldAddEndPuncttrue
\mciteSetBstMidEndSepPunct{\mcitedefaultmidpunct}
{\mcitedefaultendpunct}{\mcitedefaultseppunct}\relax
\EndOfBibitem
\bibitem[Li \emph{et~al.}(2020)Li, Zhang, Lin, and Li]{li2020pattern}
Z.-Y. Li, D.-Q. Zhang, S.-Z. Lin and B.~Li, \emph{Phys. Rev. Lett.}, 2020,
  \textbf{125}, 098002\relax
\mciteBstWouldAddEndPuncttrue
\mciteSetBstMidEndSepPunct{\mcitedefaultmidpunct}
{\mcitedefaultendpunct}{\mcitedefaultseppunct}\relax
\EndOfBibitem
\bibitem[Cross and Hohenberg(1993)]{cross1993}
M.~C. Cross and P.~C. Hohenberg, \emph{Reviews of Modern Physics}, 1993,
  \textbf{65}, 851--1112\relax
\mciteBstWouldAddEndPuncttrue
\mciteSetBstMidEndSepPunct{\mcitedefaultmidpunct}
{\mcitedefaultendpunct}{\mcitedefaultseppunct}\relax
\EndOfBibitem
\bibitem[Cross and Greenside(2009)]{cross2009}
M.~Cross and H.~Greenside, \emph{Pattern Formation and Dynamics in
  Nonequilibrium Systems}, Cambridge University Press, 2009\relax
\mciteBstWouldAddEndPuncttrue
\mciteSetBstMidEndSepPunct{\mcitedefaultmidpunct}
{\mcitedefaultendpunct}{\mcitedefaultseppunct}\relax
\EndOfBibitem
\bibitem[James \emph{et~al.}(2021)James, Suchla, Dunkel, and
  Wilczek]{james2021emergence}
M.~James, D.~A. Suchla, J.~Dunkel and M.~Wilczek, \emph{Nature communications},
  2021, \textbf{12}, 5630\relax
\mciteBstWouldAddEndPuncttrue
\mciteSetBstMidEndSepPunct{\mcitedefaultmidpunct}
{\mcitedefaultendpunct}{\mcitedefaultseppunct}\relax
\EndOfBibitem
\end{mcitethebibliography}
\bibliographystyle{rsc}

\end{document}